%% file: KK-spherical_revisited_final_rev.tex
\documentclass[11pt,fleqn]{article}

\usepackage{amsmath,amsfonts,amssymb,latexsym,cite}
\usepackage{graphicx}
\usepackage{euscript,amssymb,epsfig}
\usepackage[english]{babel}

\usepackage{color}

\newcommand{\be}[1]{\begin{equation}\label{#1}}
\newcommand{\ee}{\end{equation}}
\newcommand{\ba}[1]{\begin{eqnarray}\label{#1}}
\newcommand{\ea}{\end{eqnarray}}

\newcommand{\nn}{\nonumber}

\input preamble.tex

\begin{document}
\twocolumn [
\jnumber{}{2019}

\Title{Weak-field limit of Kaluza-Klein model with non-linear perfect fluid}
	
\Aunames{Ezgi Yal\c{c}{\i}nkaya\au{1}\foom{1}, Alexander Zhuk\au{1,}\au{2}\foom{2}}
\Addresses{\adr{1}{Department of Physics, Istanbul Technical University, 34469 Maslak, Istanbul, Turkey}\adr{2}{Astronomical Observatory, Odessa National University, Dvoryanskaya st. 2, 65082 Odessa, Ukraine}}

\Dates{}{}{}

\Abstract
{The main purpose of our paper is to construct a viable Kaluza-Klein model satisfying the observable constraints. To this end, we investigate the six-dimensional model with spherical compactification of the internal space. Background matter is considered in the form of a perfect fluid with non-linear equations of state both in the external/our and internal spaces and the model is set to include an additional bare cosmological constant $\Lambda_6$. In the weak-field approximation, the background is perturbed by pressureless gravitating mass that is a static point-like particle. The non-linearity of the equations of state of a perfect fluid makes it possible to solve simultaneously a number of problems. The demand that the parameterized post-Newtonian parameter $\gamma$ be equal to 1 in this configuration, first, ensures compatibility with gravitational tests in the Solar system (deflection of light and time delay of radar echoes) at the same level of accuracy as General Relativity. Second, it translates into the absence of internal space variations so that the gravitational potential coincides exactly with the Newtonian one, securing the absence of the fifth force. Third, the gravitating mass remains pressurless in the external space as in the standard approach to non-relativistic astrophysical objects and meanwhile, acquires effective tension in the internal space.  
}
\bigskip
	
] 
	\email{1}{ezgicanay@itu.edu.tr}
	\email{2}{ai.zhuk2@gmail.com}
\section{Introduction}

Despite the fact that the idea of the multidimensionality of spacetime is about a hundred years old (starting with pioneering works by Th. Kaluza
and O. Klein \cite{KK}), it yet remains popular in modern physical science. On the one hand, this is related to the ongoing attempts to construct theories of unification of physical interactions such as superstrings, supergravity and M-theory which have the most self-consistent formulation in spacetime with extra dimensions (see e.g. \cite{Pol}).
On the other hand, it is encouraged by the efforts to solve the problems of dark matter and dark energy as well as the hierarchy problem by introducing additional dimensions \cite{Starob,hierarchy}. Obviously, such multidimensional theories should not contradict the experimental data, in particular, the well-known gravitational tests in the Solar system: deflection of light, time delay of radar echoes and perihelion precession of Mercury.  We investigated this problem in a number of our previous papers 
\cite{our0,our1,our2,our3,our4,our5,our6,our7} and one of the main conclusions of these papers is that in the case of Ricci-flat internal spaces, the model with pressureless  gravitating mass contradicts gravitational tests \cite{our0}. 
The negative result follows as a consequence of massless fluctuations of the internal space that generate the fifth force \cite{our8}. In order to avoid these fluctuations, the gravitating mass should acquire negative pressure (i.e. tension) in the extra dimensions  \cite{our8} as common to black strings and black branes which are explicit examples of such gravitating objects \cite{our1,our2}. 
Once the tension is introduced, the model
achieves agreement with gravitational tests for the parameterized post-Newtonian (PPN) parameter $\gamma$ at the same level of accuracy as General Relativity. Nonetheless, the physical reason for negative relativistic pressure remains unclear and such requirement appears to be the main problem of this model.

In the case of spherical compactification of the internal space, the situation is handled differently; i.e., to provoke curved background, one first needs to introduce background matter \cite{our3}. For this type of models, it is possible to achieve agreement with gravitational tests in two cases. First, similar to models with toroidal compactification, we can exclude fluctuations of the internal space with the help of tension (a la blackstrings/blackbranes) \cite{our4}. Second, in the absence of tension, we may attain sufficiently short ranges of interaction for the fifth force for large enough mass of radions \cite{our3}. At first glance, it seems that the latter option well meets the necessary requirements; it is in agreement with gravitational tests and, additionally, the gravitating mass lacks tension in the internal space. However, fluctuations of background matter concentrate around the gravitating mass and the bare gravitating mass gets covered by this coat to attain effective relativistic pressure in the external space. Certainly, this contradicts the observations (e.g., Sun does not have such relativistic pressure) and the only way to avoid such problem is to introduce, again, tension in the internal space \cite{our5,our9}. 

In paper \cite{our5}, an interesting example of a model with spherical compactification was demonstrated; the bare gravitating mass without tension was shown to achieve agreement with observations while effective relativistic pressure in the external/our space was successfully avoided. The result was obtained through fine tuning between the components of a multicomponent background perfect fluid with linear equations of state. An interesting additional aspect of the model was the emergence of effective tension for the initially pressureless bare gravitating mass. This example led to the next natural question: Is it possible to build a model that has all the properties listed above, but contains a single-component background perfect fluid instead? The present paper gives an affirmative answer. Here, rather than some multicomponent background perfect fluid, we consider just one, called the non-linear perfect fluid which has non-linear equations of state in both the external and internal spaces. We demonstrate that this model has all the positive properties inherent in the multicomponent perfect fluid case and similar to the model in \cite{our5}, the fine tuning of constituent parameters appears to be the price for this result.       

The paper is structured as follows: In Section 2, we describe the background model and present metric perturbations together with pressure and energy density fluctuations of the background matter upon introduction of the gravitating source. In Section 3, first order metric corrections are determined and fine tuning conditions are imposed on these corrections for purposes of compatibility with gravitational tests. Main results and consequences of fine tuning of parameters are briefly summarized in the concluding Section 4. 

\section{Background and perturbed \\models}
 
We consider the Kaluza-Klein model with spherical compactification of the internal space. Background static metrics 
\bearr
ds^2= g^{(0)}_{ik}dx^i dx^k \nnn
= c^2dt^2-dx^2-dy^2-dz^2-a^2(d\xi^2+\sin^2\xi d\eta^2), 
\nn
\ear
is defined on a product manifold $M=M_4\times S^2$ where $M_4$ is the four-dimensional Minkowski spacetime and $S^2$ is the two-dimensional sphere of radius $a$. We also include in the model a bare multidimensional cosmological constant $\Lambda_6$. Metrics (1) should satisfy the Einstein equation
\beq
R_{ik}-\frac{1}{2}Rg_{ik}-\kappa\Lambda_6 g_{ik}=\kappa T_{ik}\, ,
\eeq
where $\kappa \equiv2S_5\tilde{G}_{6}/c^4 $ with the total solid angle $S_5=2\pi^{5/2}/\Gamma (5/2)=8\pi^2/3$ and $\tilde G_{6}$, the gravitational constant in the six-dimensional space-time.

The form of background matter can be obtained from \eq (2) after substitution of metrics (1):
\beq
T^{(0)}_{ik} = \left\{
\begin{array}{ll}
	[1/(\kappa a^2)-\Lambda_6]g^{(0)}_{ik} & \mbox{for }i,k=0,..,3; \\
	-\Lambda_6 g^{(0)}_{ik} & \mbox{for }i,k=4,5 \, .
\end{array}
\right. 
\eeq
Such energy-momentum tensor can be written in the form of a perfect fluid: 
\beq \left(T^{i}_{k}\right)^{(0)}=\diag(\bar{\varepsilon},-\bar{p}_0,-\bar{p}_0,-\bar{p}_0,-\bar{p}_1,-\bar{p}_1)\, ,
\eeq
where we introduce background parameters
\beq
\bar{\varepsilon} \equiv \frac{1}{\kappa a^2}-\Lambda_6\, ,\quad \bar{p}_0 \equiv -\left(\frac{1}{\kappa a^2}-\Lambda_6\right)\, ,\quad
\bar{p}_1\equiv \Lambda_6\, .
\eeq
In what follows, the upper bar always denotes the background values.

The energy-momentum tensor in the form (3) or (4) corresponds to the background value of a perfect fluid with non-linear equations of state in the external and internal spaces: $p(\varepsilon) = (p_0(\varepsilon),p_1(\varepsilon))$. Up to linear perturbations,  this equation of state reads
\bearr
p(\varepsilon) = \left(p_0(\varepsilon),p_1(\varepsilon)\right)\nnn
=
\left(p_0(\bar\varepsilon),p_1(\bar \varepsilon)\right)+
\left(\frac{\partial p_0}{\partial \varepsilon},
\frac{\partial p_1}{\partial \varepsilon}\right)
_{\bar\varepsilon}\delta\varepsilon + O(\delta\varepsilon^2)\nnn
\equiv \left(\bar\omega_0,\bar\omega_1\right)\bar\varepsilon+
\left(\omega_0,\omega_1\right) \delta\varepsilon
+ O(\delta\varepsilon^2)\, .
\ear
Taking into account that $\bar p_0 \equiv p_0(\bar\varepsilon)$ and $\bar p_1 \equiv p_1(\bar\varepsilon)$, \eqs (3) and (4) for the background parameters $\bar\omega_0$ and $\bar\omega_1$ yield 
\beq
\bar\omega_0=-1\, ,\quad  \bar\omega_1 = \frac{\Lambda_6}{1/\left(\kappa a^2\right)-\Lambda_6}\, . 
\eeq
New parameters $\omega_0$ and $\omega_1$ denote the first derivatives of pressures in the external and internal spaces that are calculated for the background value of the perfect fluid energy density $\bar\varepsilon$.
Therefore, pressure fluctuations in the external and internal spaces can be expressed  via fluctuations of the energy density $\delta\varepsilon$ of the perfect fluids  as follows:
\beq
\delta p_0=\omega_0\delta\varepsilon\, ,\quad \, 
\delta p_1=\omega_1\delta\varepsilon\, .
\eeq
It is worth noting that in the case of non-linear perfect fluids in general, $\bar\omega_0\neq\omega_0$ and $\bar\omega_1\neq\omega_1$.

In our model we assume that fluctuations (8) are caused by the gravitating mass in the form of a static point-like particle. This mass is pressureless in both the external and internal spaces. Accordingly, the only non-zero component of its energy density is $\hat{T}^0_0=\hat\rho c^2$, where $\hat\rho$ is the mass density of the gravitating object. In the case of uniform smearing of the gravitating mass over the internal space, the multidimensional $\hat\rho$ and three-dimensional $\hat\rho_3$ mass densities are related as $\hat\rho=\hat\rho_3/V_2$, where $V_2=4\pi a^2$ is the volume of the internal space; $\hat\rho_3 (r_3)=m\delta (\textbf{r}_3)$ and $r_3=|\textbf{r}_3| = \sqrt{x^2+y^2+z^2}$.

The total perturbed energy-momentum tensor is
\beq
T^i_k= \tilde{T}^{i}_{k}+ \hat{T}^i_k\, ,
\eeq
where 
\beq
\tilde{T}^{i}_{k} = \left\{
\begin{array}{ll}
	(\bar{\varepsilon}+\delta\varepsilon)\delta^i_k & \mbox{for }i,k=0, \\
	-(\bar{p}_0+\delta p_0)\delta^i_k & \mbox{for }i,k=1,2,3, \\
	-(\bar{p}_1+\delta p_1)\delta^i_k & \mbox{for }i,k=4,5 \, .
\end{array}
\right  . 
\eeq
 
The gravitating mass perturbs also the background metrics (1). Since the gravitating source is spherically symmetric and pressure in each factor space is isotropic, the perturbed metrics preserves its block-diagonal form (for mathematical justification of this claim, see \cite{our7}): 
\bearr
ds^2=\left[1+A^1(x,y,z)\right]c^2dt^2\nnn
-\left[1-B^1(x,y,z)\right](dx^2+dy^2+dz^2)\nnn
-\left[a^2-E^1(x,y,z)\right](d\xi^2+\sin^2\xi d\eta^2) \, . 
\ear
We suppose that $\hat\rho, \delta\varepsilon, \delta p_0, \delta p_1, A^1, B^1$ and $E^1$ are of the same order of smallness.

Our task now is to determine the first order corrections to the metric coefficients $A^1, B^1$ and $E^1$. To perform it, we should substitute the perturbed metrics (11) and perturbed energy-momentum tensor (9) in the Einstein equation
\beq R_{ik}=\kappa\left(T_{ik}-\frac{1}{4}Tg_{ik}-\frac{1}{2}\Lambda_6g_{ik}\right)\, .
\eeq

\section{Solution of perturbed \\Einstein equation} 

First, we consider the diagonal components of the Einstein equation (12).  To get the expressions for the components of the Ricci tensor, we can use the results of Appendix A in \cite{our3}.  Then, the $00$-component of (12) is
\beq
\frac{1}{2}\Delta_3A^1
=\frac{3\kappa}{4}\left[\delta\varepsilon+\delta p_0+\hat{\rho}c^2\right]+\frac{\kappa}{2}\delta p_1\, .
\eeq
Similarly, for the $11$-, $22$- and $33$- components we have:
\beq
\frac{1}{2}\Delta_3B^1
=\frac{\kappa}{4}\left[\delta\varepsilon+\delta p_0+\hat{\rho}c^2\right]-\frac{\kappa}{2}\delta p_1\, , 
\eeq
and the 44- and 55- components  read 
\bear
\frac{1}{2}\Delta_3E^1
=-\frac{1}{a^2}E^1+\frac{\kappa}{4}\left[a^2(\delta\varepsilon-3\delta p_0+\hat{\rho}c^2)\right]\nn
+\frac{\kappa}{2}a^2\delta p_1
\, . 
\ear

According to Appendices A and B in \cite{our3}, the off-diagonal components are either identically zero or, finally result in the relation
\beq
-A^1+B^1+2\frac{E^1}{a^2}=0\, .
\eeq 
Taking this relation into account, we get
\bear
\Delta_3E^1&=&\frac{a^2}{2}\left(\Delta_3A^1-\Delta_3B^1\right)\nnn
=\frac{a^2}{2}\kappa\left[\delta\varepsilon+\delta p_0+\hat{\rho}c^2+2\delta p_1\right] 
\ear
and comparing the results with the expression in (15), we obtain the connection between fluctuations $E^1$ and $\delta p_0$ as
\beq
\frac{E^1}{a^4}=-\kappa\delta p_0 \, . 
\eeq
It is important to stress that the function $E^1$ describes the variations of the internal space, i.e., the variations responsible for the appearance of the fifth force that may lead to contradiction with gravitational tests. Now, we will demonstrate how we can avoid this problem in the present model.   

Let us assume that, similar to General Relativity, the PPN parameter $\gamma=1$. To get it we should put $B^1/A^1=1$ \cite{Will}; this is our fine tuning condition. As it follows from (13) and (14), this takes place if 
\beq
\delta\varepsilon+\delta p_0+2\delta p_1+\hat{\rho}c^2=0\, ,
\eeq 
which indicates
\beq
\delta\varepsilon = \frac{-1}{1+\omega_0+2\omega_1}\hat\rho c^2\, , 
\eeq
taking into account also the expressions in (8).
For the condition (19), \eq (17) takes the form
\beq
\Delta_3E^1=0 \, ,
\eeq
with the trivial solution
\beq
E^1=0\, .
\eeq
Hence, internal space excitation does not take place (a la black string) and there is no such problem as the fifth force. As it follows from (18), this is possible for
\beq
\delta p_0=0\quad  \then \quad \omega_0=0\, ,
\eeq
which means that the background perfect fluid is in the extremum position of the equation of state in the external space: $\left(\partial p_0/\partial\varepsilon\right)_{\bar\varepsilon}=\omega_0=0$.
Then, \eq (20) reads
\beq
\delta\varepsilon = \frac{-1}{1+2\omega_1}\hat\rho c^2\, ,\quad \omega_1\neq -1/2\, .
\eeq
It can easily be seen that a particular case $\omega_1=0 \then \delta\varepsilon = -\hat\rho c^2$ results in the 
trivial model with $A^1=B^1=0$ and hence, excluded. If $\omega_1\neq 0$, then in accordance with condition (19), the gravitational potential coincides with the Newtonian one:
\beq
A^1=B^1=\frac{2\varphi_N}{c^2}\, ,\quad \varphi_N=-\frac{G_N m_{\mathrm{grav}}}{r_3}\, . 
\eeq
The gravitating mass $m_{\mathrm{grav}}$ in the above expression relates to the bare mass $m$ through
\beq
m_{\mathrm{grav}}= m \frac{2\omega_1}{1+2\omega_1} 
\eeq
and the Newtonian gravitational constant relates to its multidimensional counterpart via
\beq
4\pi G_N = \frac{S_5}{V_2}\tilde{G}_{6}\, .
\eeq
The similar relation between the Newtonian and multidimensional gravitating constants was obtained e.g. in \cite{our3}. Eq.(25) demonstrates that the condition for positiveness of the gravitating mass requires also the positiveness of the parameter $\omega_1$. 

Eventually, we get a model which, first, does not contradict the observations with respect to the PPN parameter $\gamma$ and, second, since $\delta p_0=0$, prevents the gravitating mass from picking up effective relativistic pressure in the external space. Namely, the gravitating object remains pressureless in our space as it should. For any $\omega_1 \neq -1/2, 0$; however, the gravitating mass acquires effective tension in the internal space. To show it, we should note that since $\delta\varepsilon \propto -\hat\rho c^2$, fluctuations of background matter concentrate around the gravitating mass. Then, its effective energy density reads
\beq
\delta\varepsilon_{\mathrm{eff}}= \delta\varepsilon + \hat\rho c^2
=-2\omega_1\delta\varepsilon\, ,\quad \forall \; \omega_1\neq -1/2, 0
\eeq
and for the effective equation of state parameter we get
\beq
\omega_{1\mathrm{eff}}= \frac{\delta p_1}{\delta\varepsilon_{\mathrm{eff}}}=-\frac12 <0\, ,
\eeq
which corresponds to the black strings/branes equation of state. As we already mentioned in the Introduction, the physical origin of such type of tensions for the black strings/branes is unclear. Our model proposes a possible mechanism for the occurrence of tension in the internal space for astrophysical objects which initially had negligibly small pressure (as compared to the energy density) in all dimensions.

\section{Conclusion} 

In the present paper we have considered the Kaluza-Klein model where the internal space undergoes spherical compactification. As the background matter providing such compactification, we have introduced the six-dimensional bare cosmological constant together with a perfect fluid that has non-linear equations of state both in the external and internal spaces. We have fine tuned the parameters of the model with the demand that the PPN parameter $\gamma$ be equal to 1, similar to General Relativity, which is in very good agreement with observations. In this case, the fluctuations of the internal space proved absent and a fifth force did not arise. Additionally, the gravitating mass remained pressureless in the external/our space. This is an important point since  inside astrophysical objects similar to our Sun, pressure is much less than the energy density and negligible pressure is a good approximation for the calculation of PPN parameters \cite{our0,Will,Landau}. The appearance of effective tension for the gravitating mass in the internal space stands out as a very interesting feature of this model since it presents a possible mechanism for the emerging of tension for initially tensionless gravitating objects. 

\section*{Acknowledgments} 
 The authors are grateful to Maxim Eingorn
for stimulating discussions and valuable comments.

\end{document}

%% file: preamble.tex
\usepackage{amsfonts,amssymb,cite}
\usepackage{graphicx}

\def\noi{\noindent}
\def\jnumber#1#2{\thispagestyle{empty} \noi\unitlength=1mm
    	\begin{picture}(178,10)
            \put(177,15){\llap{\large\it Grav. Cosmol. No.\,#1, #2}}
                    \end{picture}}

\newcommand{\Title}[1]{\noi {{\Large\bf #1}}\\[1ex]}

\def\Aunames#1{\noi{\bf #1}}
\def\au#1{${}^{#1}$}
\def\Addresses#1{\medskip\noi \protect
	\begin{description}\itemsep -3pt {\it #1} \end{description}}
\def\adr#1#2{\item[${}^{#1}$]{\it #2}}

\newcommand{\Dates}[3]{\noi {\it Received #1}\\
		          \noi {\it Received in final form #2}\\			
		          \noi {\it Accepted #3}\\} 
\newcommand{\Abstract}[1]{\vskip 2mm \begin{center}
        \parbox{16.4cm}{\small\noi #1} \end{center}\medskip}
\newcommand{\foom}[1]{\protect\footnotemark[#1]}

\def\email#1#2{\footnotetext[#1]{e-mail: #2}\addtocounter{footnote}{1}}

\def\nqq{\hspace*{-2em}}

\def\Jl#1#2{#1 {\bf #2},\ }

\def\ApJ#1 {\Jl{Astroph. J.}{#1}}
\def\CQG#1 {\Jl{Class. Quantum Grav.}{#1}}
\def\DAN#1 {\Jl{Dokl. AN SSSR}{#1}}
\def\GC#1 {\Jl{Grav. Cosmol.}{#1}}
\def\GRG#1 {\Jl{Gen. Rel. Grav.}{#1}}
\def\JETF#1 {\Jl{Zh. Eksp. Teor. Fiz.}{#1}}
\def\JETP#1 {\Jl{Sov. Phys. JETP}{#1}}
\def\JHEP#1 {\Jl{JHEP}{#1}}
\def\JMP#1 {\Jl{J. Math. Phys.}{#1}}
\def\NPB#1 {\Jl{Nucl. Phys. B}{#1}}
\def\NP#1 {\Jl{Nucl. Phys.}{#1}}
\def\PLA#1 {\Jl{Phys. Lett. A}{#1}}
\def\PLB#1 {\Jl{Phys. Lett. B}{#1}}
\def\PRD#1 {\Jl{Phys. Rev. D}{#1}}
\def\PRL#1 {\Jl{Phys. Rev. Lett.}{#1}}

\def\lal{&&\nqq {}}
\def\eq{Eq.\,}
\def\eqs{Eqs.\,}
\def\beq{\begin{equation}}
\def\eeq{\end{equation}}
\def\bear{\begin{eqnarray}}
\def\bearr{\begin{eqnarray} \lal}
\def\ear{\end{eqnarray}}
\def\earn{\nonumber \end{eqnarray}}
\def\nn{\nonumber\\ {}}

\def\nnn{\nonumber\\ \lal }

\def\diag{\mathop{\rm diag}\nolimits}

\def\then{\ \Rightarrow\ }



%% file: KK-spherical_revisited_final_rev.bbl
\small
\begin{thebibliography}{99}
	
	\bibitem{KK}
	Th. Kaluza, ``Zum Unit\"{a}tsproblem der Physik,'' Sitzungsber. d. Preuss. Akad. d. Wiss. 966-972 (1921);
	O. Klein, ``Quantentheorie and funfdimensionale Relativitatstheorie,'' Zeitschrift f\"{u}r Physik
	\textbf{37}, 895-906 (1926).
	
	\bibitem{Pol}
	J. Polchinski, \textit{String Theory, Volume 2: Superstring Theory and Beyond} (Cambridge University Press, Cambridge, 1998).

	\bibitem{Starob}
	U. G\"unther, A. Starobinsky, A. Zhuk,
	 ``Multidimensional cosmological models: cosmological and
		astrophysical implications and constraints,''  \PRD{69} 044003 (2004) [arXiv:hep-ph/0306191].

	\bibitem{hierarchy}
	N. Arkani-Hamed, S. Dimopoulos and G. Dvali, ``Phenomenology, astrophysics, and cosmology of theories with submillimeter dimensions and TeV scale quantum gravity,'' \PRD{59} 086004 (1999) [arXiv:hep-ph/9807344]. 
	
	\bibitem{our0}
	M. Eingorn and A. Zhuk, 
	``Classical tests of multidimensional gravity: negative result,'' \CQG{27} 205014 (2010) [arXiv:1003.5690 [gr-qc]].

	\bibitem{our1}
	 M. Eingorn and A. Zhuk, ``Kaluza-Klein models: can we construct a viable example?,'' \PRD{83} 044005 (2011) [arXiv:1010.5740 [gr-qc]]. 
	 
	 \bibitem{our2}
	  M. Eingorn , O. Medeiros, L. Crispino  and A. Zhuk,
	 ``Latent solitons, black strings, black branes, and equations of state in Kaluza-Klein models,'' \PRD{84} 024031 (2011) [arXiv:1101.3910 [gr-qc]].
	 
	 \bibitem{our3}
	 A. Chopovsky, M. Eingorn and A. Zhuk, ``Weak-field limit of Kaluza-Klein models with spherical compactification: experimental constraints,'' \PRD{85} 064028 (2012) [arXiv:1107.3388 [gr-qc]].

	\bibitem{our4}
	A. Chopovsky, M. Eingorn and A. Zhuk, ``Exact and asymptotic black
	branes with spherical compactification,'' \PRD{86} 024025 (2012) [arXiv:1202.2677 [gr-qc]].
	
	 \bibitem{our5}
	 M. Eingorn, S. H. Fakhr and A. Zhuk, ``Kaluza-Klein models with
	 spherical compactification: observational constraints and possible examples,'' \CQG{30} 115004 (2013) [arXiv:1209.4501 [gr-qc]]. 
	
	 \bibitem{our6}
	 A. Chopovsky, M. Eingorn and A. Zhuk, ``Problematic aspects of
	 Kaluza-Klein excitations in multidimensional models with Einstein internal spaces,'' \PLB{736}, 329 (2014) [arXiv:1402.1340 [gr-qc]]. 
	 
	\bibitem{our7}
	\"O. Akarsu, A. Chopovsky, A. Zhuk, ``Black branes and black strings in the astrophysical and cosmological context,''
	\PLB{778} 190-196 (2018) [arXiv:1711.08372 [gr-qc]]. 
	
	\bibitem{our8}
	 M. Eingorn and A. Zhuk, ``Remarks on gravitational interaction in
	Kaluza-Klein models,'' \PLB{713} 154 (2012) [arXiv:1201.1756 [gr-qc]]. 
	
	\bibitem{our9}
	M. Eingorn and A. Zhuk, ``Significance of tension for gravitating
	masses in Kaluza-Klein models,'' \PLB{716} 176 (2012) [arXiv:1202.4773 [gr-qc]].
	
	\bibitem{Will}
	C.M. Will, \textit{Theory and Experiment in Gravitational
	Physics} (Cambridge University Press, Cambridge,
	2000).

	\bibitem{Landau}
	L.D. Landau and E.M. Lifshitz, \textit{The Classical Theory of
	Fields, Fourth Edition: Volume 2 (Course of Theoretical
	Physics Series} (Pergamon Press, Oxford, 2000).
	
	

	
\end{thebibliography}
